\documentclass[twocolumn,aps,prl]{revtex4}
\usepackage{eurosym}
\usepackage{amsmath}
\usepackage{graphicx}
\usepackage{subfigure}
\usepackage[british]{babel}

\graphicspath{{./unconverted.eps/},{./fig.2/}}

\begin{document}

\title{Viscoelastic contact mechanics between randomly rough surfaces}
\author{M. Scaraggi$^{1,2}$ and B.N.J. Persson$^2$}
\affiliation{$^1$DII, Universit\'a del Salento, 73100 Monteroni-Lecce, Italy, EU}
\affiliation{$^2$PGI, FZ-J\"ulich, 52425 J\"ulich, Germany, EU}

\begin{abstract}
We present exact numerical results for the friction force and the contact
area for a viscoelastic solid (rubber) in sliding contact with hard,
randomly rough substrates. The rough surfaces are self-affine fractal with
roughness over several decades in length scales. We calculate the
contribution to the friction from the pulsating deformations induced by the
substrate asperities. We also calculate how the area of real contact, $%
A(v,p) $, depends on the sliding speed $v$ and on the nominal contact
pressure $p$, and we show how the contact area for any sliding speed can be
obtained from a universal master curve $A(p)$. The numerical results are
found to be in good agreement with the predictions of an analytical contact
mechanics theory.
\end{abstract}

\maketitle





Viscoelastic solids, such as rubber or gel, have many important applications
in science and technology. Rubber friction, for example, is a topic of great
practical importance e.g., for tires, syringes, wiper blades or rubber
seals, and it results from dissipative processes involving multiple
(coupled) nano- to micro- (or more) length scales, which are related to the
relaxation and diffusion dynamics of the confined polymers\cite{Grosch,Schallamach,Ludema,Ferry}
as well as to the random interaction process\cite{P1} occurring in real interfaces.
\begin{figure}[tbh]
\begin{center}
\includegraphics[width=0.46\textwidth]{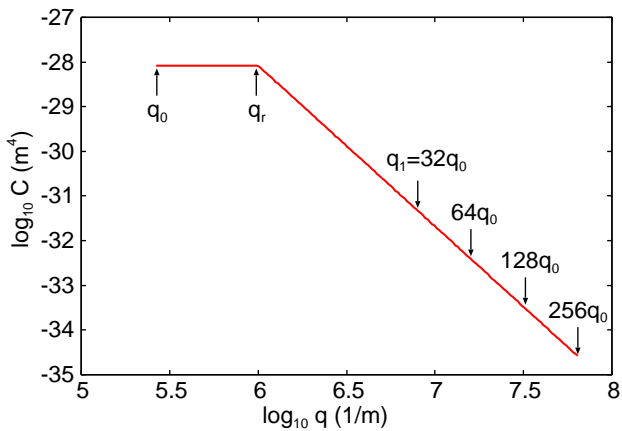}
\end{center}
\caption{Surface roughness power spectra used in the present study. The
power spectra have a low wavevector cut-off for $q_{0}=0.25\times 10^{6}\ 
\mathrm{m}^{-1}$, and a roll-off for $q_{\mathrm{r}}=4q_{0}$. For $q>q_{%
\mathrm{r}}$ the power spectra correspond to a self-affine fractal surface
with the Hurst exponent $H=0.8$. We consider 4 cases where the large
wavevector cut-off is $q_{1}=32q_{0}$, $64q_{0}$, $128q_{0}$ and $256q_{0}$;
we refer to $q_{1}=32q_{0}$ and $256q_{0}$ as the small and large system,
respectively. In the numerical calculations, a null power spectrum region is
added between $q_1$ and $\bar{q}_1=8q_1$ to improve convergence.
}
\label{Maj.Cq.Michele.ps}
\end{figure}
Due to the
(numerical) complexity of the underlying contact mechanics problem\cite{bo,carb},
involving multiple length and time scales, the friction force as well as the
real contact area between viscoelastic solids under real contact conditions
has so far only been predicted using mean field formulations of the contact
mechanics, such as the one by Persson\cite{P1} or Kl\"{u}ppel and Heinrich%
\cite{Klup}. Comparing theory with experimental results is an important
benchmark for any theory validation process, but the superposition of
coupled dissipation mechanisms encountered in rubber friction make it very
hard to test separately the different contributions to the rubber friction.
Hence, in this context any exact numerical calculations under well defined
contact characteristics, even if only possible for relatively-small systems
under idealized conditions, can furnish very useful insights into the
processes occurring in rubber sliding contacts, and test analytical theories.

In this work we make an attempt to shed light on the mechanisms of
micro-rolling friction and contact area formation in the interaction between
randomly rough surfaces of viscoelastic solids. 
In particular, we will compare the exact numerical
results with the predictions of the (more general) Persson's contact
mechanics theory. We use a recently developed residuals molecular dynamics
(RMD\cite{RMD}) scheme, adapted to the rubber viscoelastic rheology. The RMD
method has so far has been successfully applied to the investigation of
adhesive contacts between elastic solids with random roughness\cite%
{BNJMS.adhesion}, and here we extend the study to the case of sliding
contact between linear viscoelastic solids with random surface roughness.
The RMD numerical model is based on a (finite element) formulation in
wavevector space, together with a molecular dynamics modelling of the
interfacial separation in real space, the latter driven by the residuals of
the discretized contact mechanics equations. This (general-purpose) approach
allows for an equally efficient computation of the contact dynamics from
very small values of contact areas up to full contact interactions.

One of us\cite{P1} has derived a set of equations describing the friction
force acting on a rubber block sliding at the velocity $v(t)$ in contact
with a hard substrate with randomly rough surface \cite{P1}. 
For a rubber in dry contact with a hard solid with a rough surface there are
two main contributions to rubber friction, namely (\textit{i}) a
contribution derived from the energy dissipation inside the rubber due to
the pulsating deformations it is exposed to during sliding (it
could be named micro-rolling friction, as it shares the dissipation
mechanism with the classical rolling friction), and (\textit{ii}) a
contribution from the shearing processes occurring in the area of real
contact. For sliding at a constant velocity $v$, and neglecting frictional
heating, the friction coefficient due to process (\textit{i}) is: 
\begin{align}
\mu & \approx {\frac{1}{2}}\int_{q_{0}}^{q_{1}}dq\ q^{3}\ C(q)S(q)P(q)
\label{1} \\
& \times \int_{0}^{2\pi }d\phi \ \mathrm{cos}\ \phi \ \mathrm{Im}{\frac{%
E(qv\ \mathrm{cos}\ \phi )}{(1-\nu ^{2})\sigma _{0}}},  \notag
\end{align}%
where $\sigma _{0}$ is the nominal contact stress, $C(q)$ the surface
roughness power spectrum and $E(\omega )$ the rubber viscoelastic
modulus. The function $P(q)=A(\zeta )/A_{0}$ is the relative contact area
when the interface is observed at the magnification $\zeta =q/q_{0}$, where $%
q_{0}$ is the smallest (relevant) roughness wavevector. We have 
\begin{equation}
P\left( q\right) ={\frac{2}{\pi }}\int_{0}^{\infty }dx\ {\frac{\mathrm{sin}x%
}{x}}\mathrm{exp}\left[ -x^{2}G(q)\right] =\mathrm{erf}\left( {\frac{1}{%
2\surd G}}\right)  \label{P(q)}
\end{equation}%
where 
\begin{equation}
G\left( q\right) ={\frac{1}{8}}\int_{q_{0}}^{q}dq\ q^{3}C\left( q\right)
\int_{0}^{2\pi }d\phi \ \left\vert {\frac{E\left( qv\ \mathrm{cos}\ \phi
\right) }{\left( 1-\nu ^{2}\right) \sigma _{0}}}\right\vert ^{2}
\label{G(q)}
\end{equation}%
The factor $S(q)$ in (\ref{1}) is a correction factor which takes into
account that the asperity-induced deformations of the rubber is smaller than
would be in the case if complete contact would occur in the (apparent)
contact areas observed at the magnification $\zeta =q/q_{0}$. For contact
between elastic solids this factor reduces the elastic asperity-induced
deformation energy, and including this factor gives a distribution of
interfacial separation in good agreement with experiment and exact numerical
studies\cite{Alm}. The interfacial separation describes how an elastic (or
viscoelastic) solid deforms and penetrates into the roughness valleys, and
it is stressed here that these (time-dependent) deformations cause the
viscoelastic contribution to rubber friction. We assume that the same $S(q)$
reduction factor as found for elastic contact is valid also for sliding
contact involving viscoelastic solids. For elastic solids it has been found
that $S(q)$ is well approximated by 
\begin{equation*}
S(q)=\gamma +(1-\gamma )P^{2}(q),
\end{equation*}%
where $\gamma \approx 1/2$, and here we use the same expression for
viscoelastic solids, being in nature a geometrical parameter. Note that $%
S\rightarrow 1$ as $P\rightarrow 1$ which is an exact result for complete
contact. In fact, for complete contact the expression (\ref{1}) is exact
(see below). Note finally that in the original rubber friction theory\cite%
{P1} the correction factor $S(q)$ was not included.

The second contribution (\textit{ii}) to the rubber friction force,
associated with the area of (apparent) contact observed at the magnification 
$\zeta_1 = q_1/q_0$, is given by $\tau_{\mathrm{f}} A_1$. Here, $\tau_{%
\mathrm{f}} (v)$ is the (weakly) velocity-dependent effective frictional
shear stress acting in the contact area $A_1 = A(\zeta_1 ) =P(q_1) A_0$. In
this study we only consider the viscoelastic contribution to the rubber
friction, but we also study the area of real contact which is needed when
calculating the second contribution to the rubber friction.
%
%
\begin{figure}[tbh]
\begin{center}
\includegraphics[width=0.46\textwidth,angle=0]{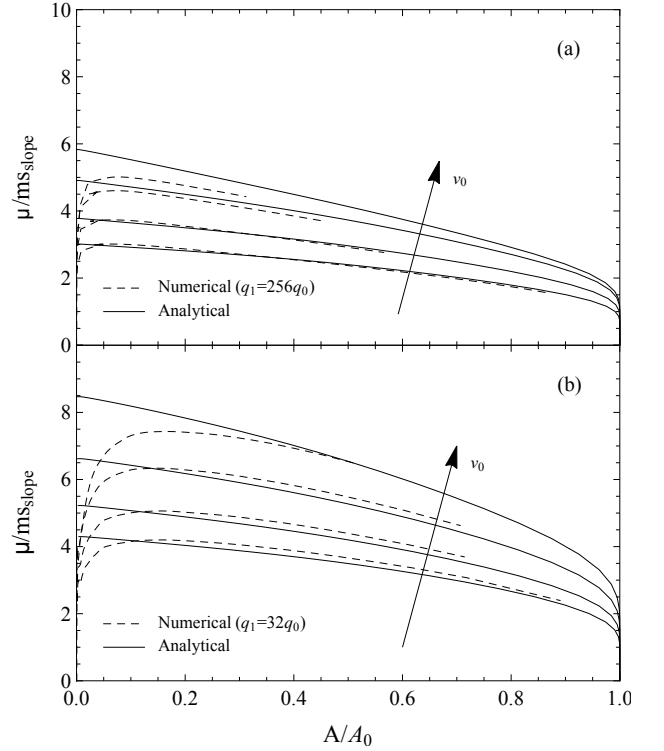}
\end{center}
\caption{The friction coefficient, divided by the ms slope, as a
function of the (normalized) area of contact $A/A_{0}$, for several sliding
speeds: $v=0.01$, $0.1$, $1$, $10\ \mathrm{m/s}$. The solid lines are the
theory predictions, while the dotted lines are from the exact numerical
study. for \textit{(a)} large system, and \textit{(b)} small system.}
\label{A.mu.eps}
\end{figure}

Note that the surface mean square slope is given by 
\begin{equation*}
\left\langle \left( \nabla h\right) ^{2}\right\rangle =2\pi
\int_{q_{0}}^{q_{1}}dq\ q^{3}C\left( q\right)
\end{equation*}%
so we can write 
\begin{gather}
{\frac{\mu }{\left\langle \left( \nabla h\right) ^{2}\right\rangle }}\approx
\label{2} \\
{\frac{\int_{q_{0}}^{q_{1}}dq\ q^{3}C\left( q\right) S\left( q\right)
P\left( q\right) \int_{0}^{2\pi }d\phi \ \mathrm{cos}\ \phi \ \mathrm{Im}{%
\frac{E\left( qv\ \mathrm{cos}\phi ,T_{0}\right) }{\left( 1-\nu ^{2}\right)
\sigma _{0}}}}{4\pi \int_{q_{0}}^{q_{1}}dq\ q^{3}C\left( q\right) }}.  \notag
\end{gather}%
For complete contact $S(q)=P(q)=1$ and if $\mathrm{Im}E(\omega
,T_{0})$ would be weakly dependent on $\omega $, the integral over $\phi $
in (\ref{2}) would be weakly dependent on $q$, and in this limiting case the
viscoelastic friction coefficient would be nearly proportional to the mean
square slope. However, these assumptions never holds in practice and the
friction coefficient cannot be simply related to the mean surface slope.%
%
\begin{figure*}[tbh]
\begin{center}
\leavevmode%
\subfigure[]  {
\includegraphics[width=0.35\textwidth,angle=0]{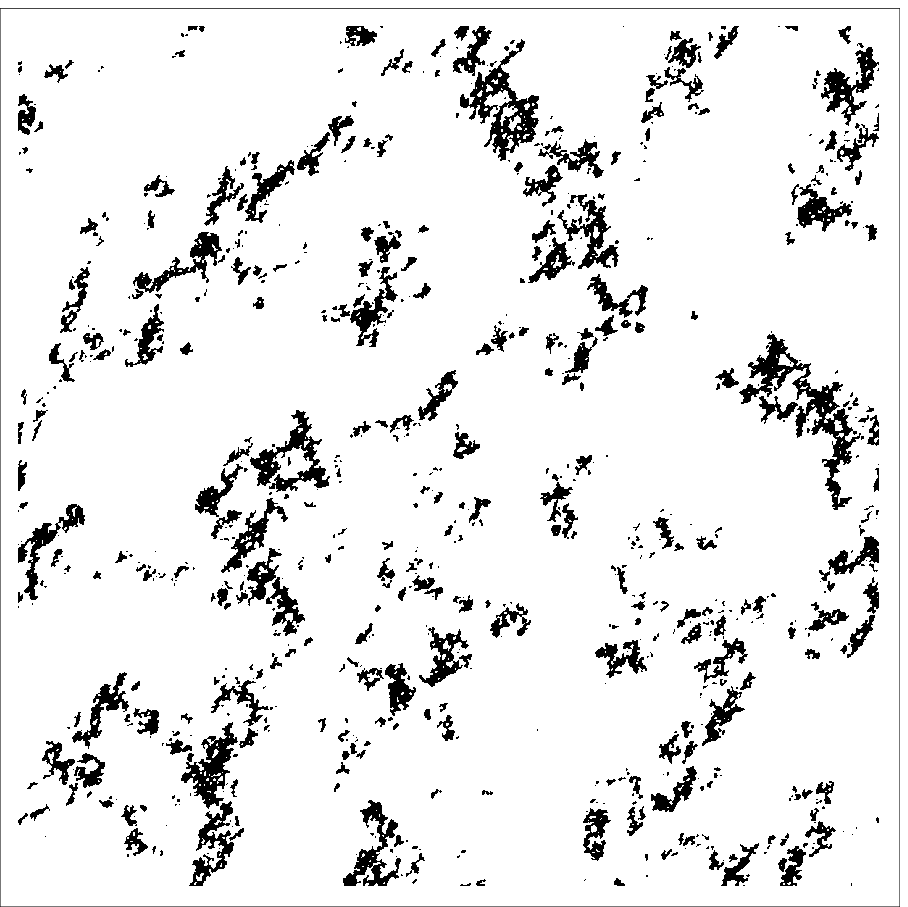}  \label{Majo.contact.large.eps}
} \qquad 
\subfigure[]  {
\includegraphics[width=0.35\textwidth,angle=0]{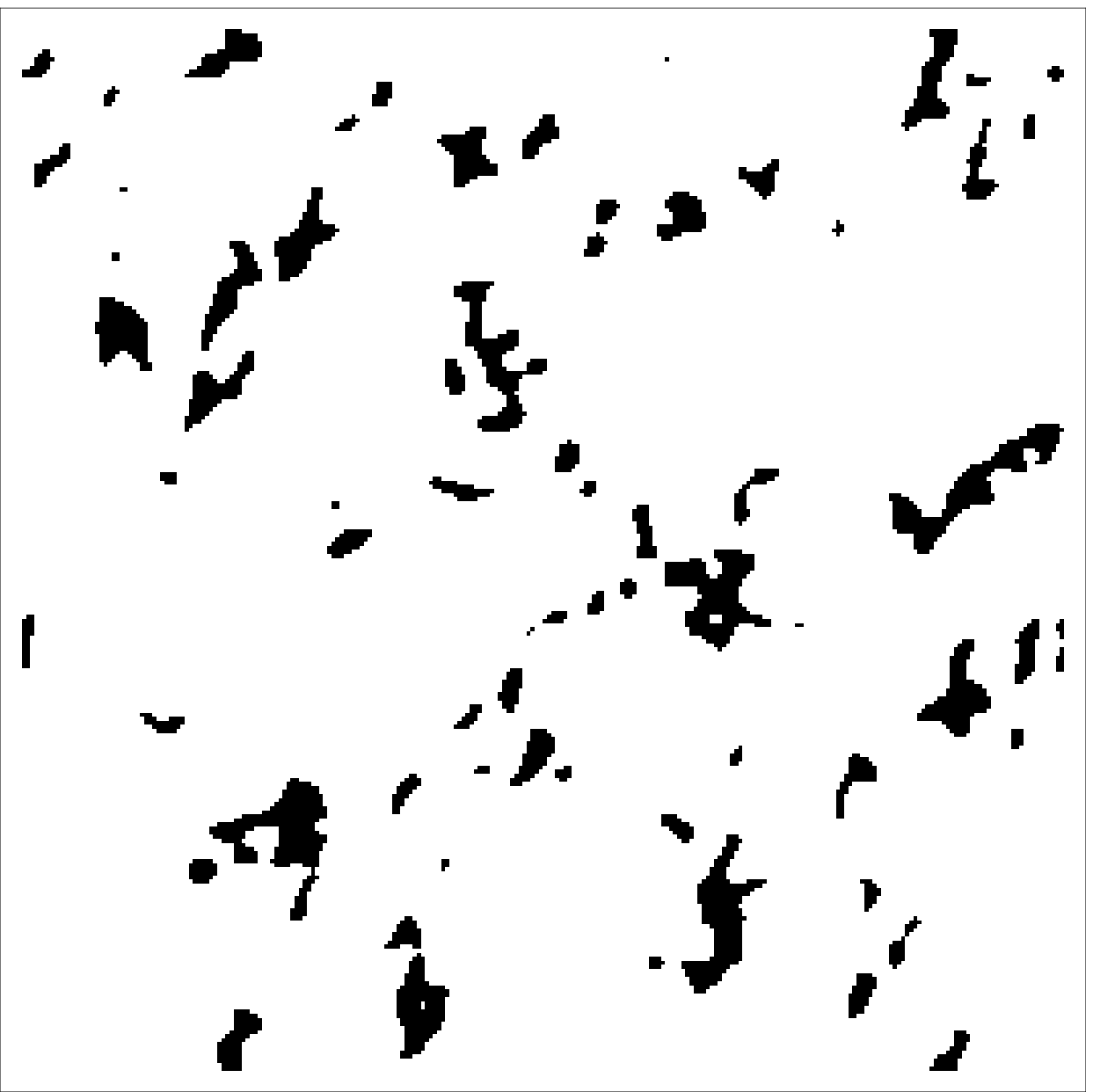}  \label{Majo.contact.small.eps}
}
\end{center}
\caption{The contact morphology for the sliding speed $v=10\ \mathrm{m/s}$
and for such a normal load that $A/A_{0}\approx 0.05$. \textit{(a)} large
system, \textit{(b)} small system. Black domains correspond to the true
contact area.}
\label{Majo.contact}
\end{figure*}
\begin{figure}[tbp]
\begin{center}
\includegraphics[width=0.46\textwidth,angle=0]{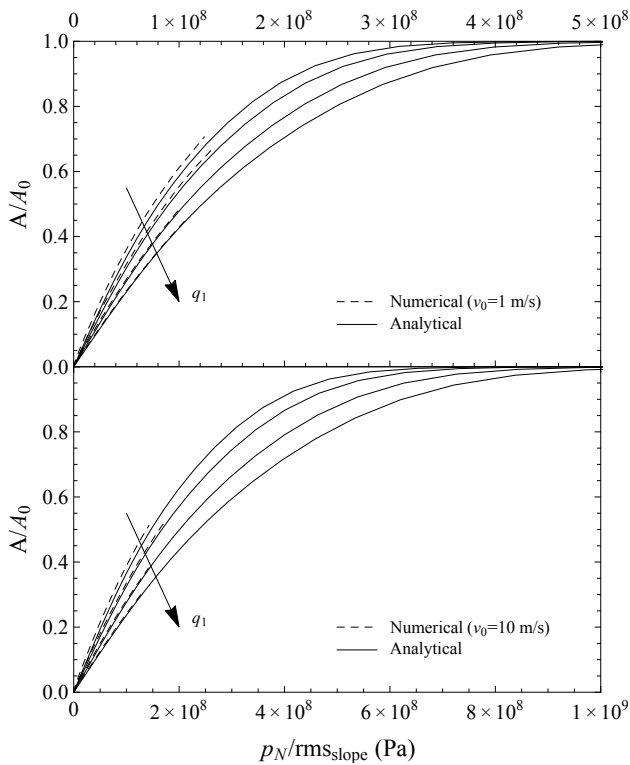}
\end{center}
\caption{The normalized area of contact $A/A_{0}$ as a function of the
contact pressure $p_{\mathrm{N}}$ (divided by the rms slope), for several
large cut-off frequencies $q_{1}$. The reported values
correspond to sliding velocities occurring in the rubbery-to-glassy 
rubber transition regime.
}
\label{DIM.all.p.A.eps}
\end{figure}
\begin{figure}[tb!]
\includegraphics[width=0.46\textwidth]{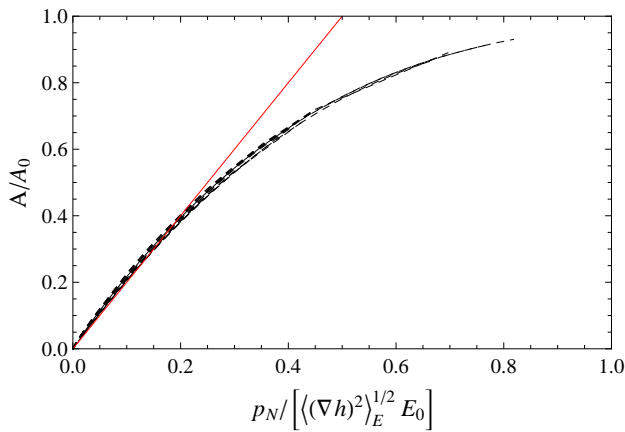}
\caption{The numerically-calculated nominal contact area $A/A_{0}$ as a
function of the contact pressure $p_{\mathrm{N}}/\left[ \left\langle \left( \protect%
\nabla h\right) ^{2}\right\rangle _{E}^{1/2} E_0\right]$ [where $\left\langle \left( 
\protect\nabla h\right) ^{2}\right\rangle _{E}^{1/2}$ is an effective
mean-square surface slope defined in (\protect\ref{effective.ms})], for
several values of roughness cut-off wavevectors ($q_{1}=32q_{0}$, $64q_{0}$, 
$128q_{0}$ and $256q_{0}$) and sliding velocities ($v=0.01$, $0.1$, $1$, $%
10\ \mathrm{m/s}$). All the curves appear superposed to an unique
mastercurve. The red line has a slope of $2$.}
\label{good.scaling.eps}
\end{figure}

We now present exact numerical results for the viscoelastic contribution to
the friction for a wide range of contact conditions, including nominal
contact pressure, sliding velocity, and the large wavevector cut-off $q_1$ (which
determines the length scales over which the surface exhibit roughness). The
RMD numerical results will be compared with the predictions of the rubber
friction theory presented above.

In the calculations we use the viscoelastic modulus $E(\omega )$ measured
for a tread rubber compound\cite{Boris}, and the substrate is assumed rigid
with a randomly rough isotropic surface. Fig. \ref{Maj.Cq.Michele.ps} shows
the surface roughness power spectra used in the present study. The power
spectra have a low wavevector cut-off for $q_{0}=0.25\times 10^{6}\ \mathrm{m%
}^{-1}$, and a roll-off for $q_{\mathrm{r}}=4q_{0}$. For $q>q_{\mathrm{r}}$
the power spectra correspond to a self-affine fractal surfaces with the
Hurst exponent $H=0.8$. We consider four cases where the large wavevector
cut-off is $q_{1}=32q_{0}$, $64q_{0}$, $128q_{0}$ and $256q_{0}$; we refer
to $q_{1}=32q_{0}$ and $256q_{0}$ as the small and large system,
respectively. The root mean square roughness is determined mainly by the
long-wavelength roughness and is therefore nearly the same for all the
different cases, with $h_{\mathrm{rms}}\approx 27\ \mathrm{nm}$.
%

Figures \ref{A.mu.eps}a and \ref%
{A.mu.eps}b show the friction coefficients [divided by
the mean square (ms) slope], as a function of the (normalized) area of contact $%
A/A_{0}$, for the large and small systems, respectively, and for several
sliding speeds: $v=0.01$, $0.1$, $1$, $10\ \mathrm{m/s}$. The solid lines
are the predictions of the Persson's rubber friction theory, whereas the
dotted lines are from the exact numerical study. Note that because of the
Hertzian-like contact for small load, the numerical friction curves show a
non-monotonic behaviour, where friction increases at small increasing values
of contact areas. For high enough loads the friction coefficient decreases
with increasing load (corresponding to increasing contact area) and the
numerical results smoothly converge to the mean field predictions. Note that
at $A/A_{0}\approx 0.05$, the small system is still in the Hertzian friction
regime\cite{GT}, whereas the large system is experiencing the transition. Fig. \ref%
{Majo.contact.large.eps} and \ref{Majo.contact.small.eps} show the contact
morphology for $A/A_{0}\approx 0.05$ for both the large and small systems,
respectively, at sliding speed $v_{0}=10\ \mathrm{m/s}$. For the large
system the contact is split in a huge number of smaller patches (compared to
the small system). When the surface exhibits roughness at shorter and
shorter length scales (i.e. when the cut-off $q_{1}$ increases) the
Hertzian-like contact will prevail only at lower and lower nominal contact
pressure, i.e. the finite size effect will be confined at smaller nominal
contact areas and the
system will move toward the thermodynamic limit, where a remarkably good
agreement with the mean field theory exists.

Fig. \ref{DIM.all.p.A.eps} shows the (normalized) area of real contact as a
function of the applied pressure $p_{\mathrm{N}}$ [divided by
the root mean square (rms) slope], for several values of
roughness cut-off wavevectors ($q_{1}=32q_{0}$, $64q_{0}$, $128q_{0}$ and $%
256q_{0}$) and sliding velocities ($v=1$, $10\ \mathrm{m/s}$%
). In particular, the solid lines are the theory results and the dotted
lines are from the numerical simulations. The agreement is very good. It is
observed that when the sliding velocity increases, the asperity deformation
frequencies increase and the rubber becomes elastically stiffer, resulting
in the decrease of the contact area with increasing sliding speed. Moreover,
this local stiffening depends on the perturbing frequencies which increases
when more short-wavelength roughness is added to the surface profile, i.e.,
when $q_{1}$ increases. Therefore, for viscoelastic contacts one cannot
expect the area of real contact to be proportional to the inverse of the the
root-mean-square roughness as observed for elastic contacts. This is
confirmed by Fig. \ref{DIM.all.p.A.eps} which shows that for each velocity
value, the numerically-predicted curves are not superposing, in agreement
with the analytical results (\ref{P(q)}).

However, the analytical theory (\ref{P(q)}) and (\ref{G(q)}) suggests a
possible mechanism to interpret the contact area results. As shown in Eq. (%
\ref{G(q)}), this local (scale dependent) rubber stiffening is equivalent to
an apparent increase (with respect to the static contact) of the roughness
power spectral content by a factor%
\begin{equation}
s\left( q,v_{0}\right) ={\frac{1}{2\pi }}\int_{0}^{2\pi }d\phi \ \left\vert {%
\frac{E\left( qv\ \mathrm{cos}\phi ,T_{0}\right) }{E_{0}}}\right\vert ^{2},
\label{s(q)}
\end{equation}%
where $E_{0}$ is the low frequency rubber elastic modulus [$E_{0}=E(\omega
\rightarrow 0)$]. Hence, it is now easy to define a new
(viscoelastic-dependent) effective mean square slope as 
\begin{equation}
\left\langle \left( \nabla h\right) ^{2}\right\rangle _{E}=2\pi
\int_{q_{0}}^{q_{1}}dq\ q^{3}C\left( q\right) s\left( q,v\right) ,
\label{effective.ms}
\end{equation}%
which is depending on the sliding velocity $v$ via the dependency of $%
E(\omega )$ on $\omega =qv\mathrm{cos}\phi $. In Fig. \ref{good.scaling.eps}
we shown the normalized contact area as a function of the contact pressure
scaled by the effective root mean square slope, for several values of
roughness cut-off wavevectors ($q_{1}=32q_{0}$, $64q_{0}$, $128q_{0}$ and $%
256q_{0}$) and sliding velocities ($v=0.01$, $0.1$, $1$, $10\ \mathrm{m/s}$%
). Remarkably, the theory-suggested scaling allows to obtain an unique
contact mastercurve similar to the case of purely elastic interactions.
Hence, it is recognizable that an universal scaling rules the asperity
mediated multiscale interaction of randomly rough surfaces, which is
insensitive to the particular rheological description of the bulk dynamics.

To summarize, we have performed exact numerical calculations for the
viscoelastic contribution to rubber friction, and compared the results with
the prediction of an analytical theory. Both the friction coefficient and
the area of contact are rather well described by the theory, in particular
for large contact pressure (in the limit of complete contact, the analytical
theory is exact). Viscoelasticity will introduce some anisotropy in the
contact morphology, but this effect seems to be rather unimportant for the
variation of the rubber friction and contact area with sliding speed. In the
numerical calculations we have neglected the effect of frictional heating
and strain softening, which are likely to be important in most practical
applications. These effects can be approximately included in the analytical
theory, but including the same effects in the numerically exact treatment
seams highly non-trivial.

\vskip 0.3cm \textbf{Acknowledgments} MS acknowledges FZJ for the support
and the kind hospitality received during his visit to the PGI-1, where this
work was initiated. MS also acknowledges COST Action MP1303 for grant
STSM-MP1303-090314-042252.


\begin{thebibliography}{99}

\bibitem{Grosch} K.A. Grosch, Proc. R. Soc. London, Ser. \textbf{A}274, 21 (1963)

\bibitem{Schallamach} A. Schallamach, Wear \textbf{6}(5), 375-382 (1963)

\bibitem{Ludema} K.C. Ludema, D. Tabor, Wear \textbf{9}(5), 329-348 (1966)

\bibitem{Ferry} J.D. Ferry, {\it Viscoelastic properties of polymers}, 3rd edn. (Wiley, 1980)

\bibitem{P1} B.N.J. Persson, J. Chem. Phys. \textbf{115}, 3840 (2001)

\bibitem{bo} B.N.J. Persson, O. Albohr, U. Tartaglino, A.I. Volokitin, E. Tosatti, J. Phys.: Condens. Matter \textbf{17}(1), R1-R62 (2005)

\bibitem{carb} G. Carbone, C. Putignano, Phys. Rev. E \textbf{89}(3), 032408 (2014)

\bibitem{Klup} M. Kl\"uppel and G. Heinrich, Rubber Chem. Technol. \textbf{73%
}, 578 (2000)

\bibitem{RMD} M. Scaraggi, \textit{in preparation} (2014)

\bibitem{BNJMS.adhesion} B.N.J. Persson, M. Scaraggi, \textit{submitted},
arXiv:1405.3123 [cond-mat.soft], (2014)


\bibitem{Alm} A. Almqvist, C. Campana, N. Prodanov, B.N.J Persson, J. Mech. Phys. Solids \textbf{59}, 2355 (2011)

\bibitem{Boris} B. Lorenz and B.N.J. Persson (unpublished)

\bibitem{GT} J. Greenwood, D. Tabor, J. Phys. Soc. \textbf{71}, 989 (1958)

%
%
%
%
%
%
%
%
%
%
%
%
%
%
%
%
%
%
%
%
%
%
%

\end{thebibliography}
\end{document}